
\font\BBBig=cmr10 scaled\magstep3


\def\foot#1{
\footnote{($^{\the\foo}$)}{#1}\advance\foo by 1
} 


\def\title{
{\bf\BBBig
\centerline{Supersymmetry of the magnetic vortex}
}}

\def\runningtitle{
Supersymmetry of the vortex
} 


\def\authors{
\centerline{
C.~DUVAL \foot{
Centre de Physique Th\'eorique, CNRS, 
Luminy, Case 907. 
F-13288 Marseille Cedex 9 (France). 
UMR 6207 du CNRS associ\'ee aux 
Universit\'es d'Aix-Marseille I et II et Universit\'e du Sud Toulon-Var; Laboratoire 
affili\'e \`a la FRUMAM-FR2291. mailto: duval@cpt.univ-mrs.fr.}
\quad\hbox{and}\quad
P.~A.~HORV\'ATHY\foot{
Laboratoire de Math\'ematiques et de Physique Th\'eorique,
Universit\'e de Tours, Parc de Grandmont,
F-37200 Tours (Fr). mailto: horvathy@lmpt.univ-tours.fr.
}}
}

\def\runningauthors{
Christian DUVAL and P\'eter HORV\'ATHY
} 


\voffset = 1cm 
\baselineskip = 16pt 

\headline ={
\ifnum\pageno=1\hfill
\else\ifodd\pageno\hfil\tenit\runningtitle\hfil\tenrm\folio
\else\tenrm\folio\hfil\tenit\runningauthors\hfil
\fi\fi
} 

\nopagenumbers
\footline = {\hfil} 



\font\tenb=cmmib10 
\newfam\bsfam

\textfont\bsfam=\tenb

\mathchardef\betab="080C
\mathchardef\xib="0818
\mathchardef\omegab="0821
\mathchardef\deltab="080E
\mathchardef\epsilonb="080F
\mathchardef\pib="0819
\mathchardef\sigmab="081B
\mathchardef\bfalpha="080B
\mathchardef\bfbeta="080C
\mathchardef\bfgamma="080D
\mathchardef\bfomega="0821
\mathchardef\zetab="0810


\def\parag{\hfil\break}
\def\and{\qquad\hbox{and}\qquad}

\def\kikezd{\parag\underbar}

\def\osp{\math{osp}}

\def\o{\math{o}}

\def\br{{\bf r}}
\def\bp{{\bf p}}
\def\bq{{\bf q}}

\def\bA{{\bf A}}
\def\smallcirc{{\raise 0.5pt \hbox{$\scriptstyle\circ$}}}
\def\smallover#1/#2{\hbox{$\textstyle{#1\over#2}$}}
\def\2{{\smallover 1/2}}

\def\semidirectproduct{
{\ooalign
{\hfil\raise.07ex\hbox{s}\hfil\crcr\mathhexbox20D}}}
\def\ccr{\cr\noalign{\medskip}}
\def\={\!=\!}

\def\D{D\mkern-2mu\llap{{\raise+0.5pt\hbox{\big/}}}\mkern+2mu}


\newcount\ch 
\newcount\eq 
\newcount\foo 
\newcount\ref 

\def\chapter#1{
\goodbreak\parag\eq = 1\advance\ch by 1{\bf\the\ch.\enskip#1}
}

\def\equation{
\leqno(
\the\eq)\global\advance\eq by 1
}

\def\reference{
\parag ${}^{\number\ref}$ \advance\ref by 1
}

\eq=1
\ch = 0 
\foo = 1 
\ref = 1 


\def\math#1{\mathop{\rm #1}\nolimits}


\title
\vskip 1.0cm
\authors

\vskip 0.2in

\noindent
Abstract. {\it The $N\=2$
supersymmetry of the Pauli Hamiltonian
in any static magnetic field in the plane
combines, for the magnetic vortex, with
Jackiw's bosonic $\o(2)\times\o(2,1)$ symmetry, into an
$\o(2)\times\osp(1/2)$ dynamical supersymmetry.}

\vskip3mm
\noindent{May 1993}. Tours Preprint no 60/93.
\vskip3mm

A few years ago, Jackiw [1] pointed out that a spin-$0$
particle in a Dirac monopole field has an $\o(2,1)$ dynamical
symmetry, generated by the spin-$0$ Hamiltonian, $H_0\=\pib^2/2m$
where $\pib\=\bp-e{\bf A}$, by the dilatation and by the expansion,
$$
D\=tH_0-\smallover1/4\{\pib,\br\}
\and
K\=-t^2H_0+2tD+\smallover{m}/{2}\br^2,
\equation
$$ to which angular momentum adds an $\o(3)$.
This allowed him to calculate the spectrum and the wave functions
group-theoretically [1].

Jackiw's result was extended to spin-$\2$ particles by D'Hoker and
Vinet [2] who have shown that, for the Pauli Hamiltonian
$H\=(1/2m)\left[\pib^2-e{\bf B}\cdot\sigmab\right]$,
not only the conformal generators $D$ and $K$, but also
the fermionic generators
$Q\={1/\sqrt{2m}}\,\pib\cdot\sigmab$
and
$S\=\sqrt{m/2}\,\br\cdot\sigmab-tQ$
are conserved.
Thus, the spin system admits an
$\o(3)\times\osp(1/1)$ conformal supersymmetry, yielding now
an algebraic solution of the Pauli equation [2].

More recently, Jackiw [3] found that the $\o(2,1)$ symmetry,
generated by --- formally --- the same $D$ and $K$ as
above, is also present for a magnetic vortex (an idealization for
the Aharonov-Bohm experiment [4]), allowing for a group-theoretic
treatment of the problem.

In this Letter we show that
the
$N\=2$ supersymmetry of the Pauli Hamiltonian of a
spin-$\2$ particle (present for any magnetic field in the plane [5])
combines,
 for a magnetic vortex, with
Jackiw's $\o(2)\times\o(2,1)$ into an $\o(2)\times\osp(1/2)$
superalgebra. This result is to be compared with the Galilean supersymmetry
discovered recently by Leblanc et al. [6] for non-relativistic
Chern-Simons systems, and with the $\osp(1/2)$ found by Hughes et al. 
for a constant magnetic 
field [7].

Thus the planar system has more symmetries as its higher-dimensional 
counterpart. We call this supersymmetry
exotic, because it is realized with two --- rather then four-component 
--- objects.
 Our pseudoclassical calculations in Ref. [6] 
indicate that such an \lq exotic' supersymmetry is only possible in two 
spatial dimensions --- one more indication of the particular status of 
two-dimensional physics.

Let us start with a spin-$\2$ particle in a static magnetic field
${\bf B}\=\big(0,0,B(x,y)\big)$.
Dropping the irrelevant $z$ variable, we can work in the plane.
Then our model is described by the Pauli Hamiltonian
$$
H={1\over2m}\left[\pib^2-eB\sigma_3\right],
\equation
$$
where $B\={\rm rot}\,\bA\!\equiv\!\epsilon^{ij}\partial_iA_j$ is the
scalar magnetic field.
It is now easy to see that the Hamiltonian  is a perfect
square in two different ways:
both operators\foot{The cross product of two planar vectors,
${\bf u}\times{\bf v}=\epsilon_{ij}u^iv^j$, is a scalar.}
$$
Q={1\over\sqrt{2m}}\,\pib\cdot\sigmab
\and
Q^*={1\over\sqrt{2m}}\,\pib\times\sigmab,
\equation
$$
where $\sigmab\=(\sigma_1,\sigma_2)$, satisfy
$$
\{Q,Q\}=\{Q^\star,Q^\star\}=2H.
\equation
$$
Thus, for
 any static, purely magnetic field in the plane, $H$ is an
$N\=2$ supersymmetric Hamiltonian. The supercharge $Q$ is a standard object
 used in
supersymmetric quantum mechanics [5];  the `twisted' charge $Q^\star$
was used, e.g., by Jackiw [8], to describe the Landau states
in a constant magnetic field --- a classic example of 
supersymmetric quantum mechanics [5,7]. 

Let us assume henceforth that $B$ is the field of a point-like
magnetic vortex directed along the $z$-axis, $B\=\Phi\,\delta(\br)$,
where $\Phi$ is the total magnetic flux.
This can be viewed as an idealization of the spinning version 
of the Aharonov-Bohm
experiment [9].

Inserting 
$A_i(\br)\=-(\Phi/2\pi)\,\epsilon_{ij}\,\br^j/r^2$ into the Pauli
Hamiltonian $H$, it is  straightforward to check
that
$
D=tH-\smallover1/4 \left\{\pib,\br\right\}
$
and
$K=-t^2H+2tD+\2mr^2
$
generate, along with $H$,
the $\o(2,1)$ Lie algebra:
$[D,H]\=-iH$,
$[D,K]\=iK$,
$[H,K]\=2iD$.
The angular momentum,
$J\=\br\times\pib$, adds to this $\o(2,1)$ an extra $\o(2)$.
(The correct definition of angular momentum requires boundary
conditions, see [10]).

Commuting $Q$ and $Q^\star$ with the expansion, $K$,
yields two more generators, namely
$$
S=i[Q,K]
=\sqrt{m\over2}\left(\br-{\pib\over m}t\right)\cdot\sigmab,
\qquad
S^\star
=i[Q^\star,K]
=\sqrt{m\over2}\left(\br-{\pib\over m}t\right)\times\sigmab.
\equation
$$

It is now straightforward to see that both sets $Q,S$ and
$Q^\star,S^\star$ extend
the $\o(2,1)\cong\osp(1/0)$ into an $\osp(1/1)$ superalgebra.
However, these two algebras do not close yet: the `mixed'
anticommutators $\{Q,S^\star\}$ and $\{Q^\star,S\}$ bring in a new
conserved charge, {\it viz.}
$\{Q,S^\star\}\=-\{Q^\star,S\}=J+2\Sigma$,
where $\Sigma=\2\sigma_3$.
But $J$ satisfies now non-trivial
commutation relations with the supercharges,
$[J,Q]=-iQ^\star$,
$[J,Q^\star]=iQ$,
$[J,S]=-iS^\star$,
$[J,S^\star]=iS$.
Thus, setting
$Y\=J+2\Sigma=\br\times\pib+\sigma_3$,
the generators $H,D,K,Y$ and
$Q,Q^\star,S,S^\star$ satisfy
$$
\matrix{
[Q,D]\hfill&=&\smallover i/2 Q,\hfill
&[Q^\star,D]\hfill&=&\smallover i/2 Q^\star,\hfill
\ccr
[Q,K]\hfill&=&-iS,\hfill
&[Q^\star,K]\hfill&=&-iS^\star,\hfill
\ccr
[Q,H]\hfill&=&0,\hfill
&[Q^\star,H]\hfill&=&0,\hfill
\ccr
[Q,Y]\hfill&=&-iQ^\star,\hfill
&[Q^\star,Y]\hfill&=&iQ,\hfill
\ccr
[S,D]\hfill&=&-\smallover i/2 S,\hfill
&[S^\star,D]\hfill&=&-\smallover i/2 S^\star,\hfill
\ccr
[S,K]\hfill&=&0,\hfill
&[S^\star,K]\hfill&=&0,\hfill
\ccr
[S,H]\hfill&=&iQ,\qquad\hfill
&[S^\star,H]\hfill&=&iQ^\star,\hfill
\ccr
[S,Y]\hfill&=&-iS^\star,\hfill
&[S^\star,Y]\hfill&=&iS,\hfill
\ccr
\{Q,Q\}\hfill&=&2H,\qquad\qquad\hfill
&\{Q^\star,Q^\star\}\hfill&=&2H,\hfill
\ccr
\{S,S\}\hfill&=&2K,\hfill
&\{S^\star,S^\star\}\hfill&=&2K,\hfill
\ccr
\{Q,Q^\star\}\hfill&=&0,\hfill
&\{S,S^\star\}\hfill&=&0,\hfill
\ccr
\{Q,S\}\hfill&=&-2D,\hfill
&\{Q^\star,S^\star\}\hfill&=&-2D,\hfill
\ccr
\{Q,S^\star\}\hfill&=&Y,\hfill
&\{Q^\star,S\}\hfill&=&-Y.\hfill
\cr}
\equation
$$
When added to the $\o(2,1)$ relations, this means that our generators
span the $\osp(1/2)$ superalgebra [2].
On the other hand, $Z\=J+\Sigma\=\br\times\pib+\2\sigma_3$
commutes with all generators of $\osp(1/2)$, so that the full
symmetry is the direct product
$\osp(1/2)\times\math{o}(2)$, generated by
$$
\left\{
\matrix{
Y\hfill&=&
\br\times\pib+\sigma_3,\qquad\qquad\hfill
&Q\hfill
&=&\displaystyle{1\over\sqrt{2m}}\,\pib\cdot\sigmab,\hfill
\ccr
H\hfill&=&
\displaystyle{1\over2m}\,
\left[\pib^2-eB\sigma_3\right],\qquad\quad\hfill
&Q^\star\hfill &=&
\displaystyle{1\over\sqrt{2m}}\,\pib\times\sigmab,\hfill
\ccr
D\hfill &=&
-\smallover 1/4\,
\left\{\pib,\bq\right\}
-t\displaystyle{eB\over2m}\,\sigma_3,\quad\hfill
&S\hfill
&=&\sqrt{\displaystyle{m\over2}}\,\bq\cdot\sigmab,\hfill
\ccr
K\hfill&=&
\2m\bq^2,\hfill
&S^\star\hfill&=&
\sqrt{\displaystyle{m\over2}}\,\bq\times\sigmab,\hfill
\ccr
Z\hfill&=&
\br\times\pib+\2\sigma_3,\hfill&
\cr
}\right.
\equation
$$
where we have put
$\bq\=\br\-(\pib/m)t$.

The
supersymmetric Hamiltonian (1) is the square of Jackiw's [8]
two-dimensional Dirac operator $\pib\times\sigma$.
But the Dirac operator
is supersymmetric in any even dimensional space. 
The energy levels are therefore non-negative; 
eigenstates with non-zero energy are doubly
degenerate; the system has ${\rm Ent}(e\Phi-1)$ zero-modes
[8, 9]. The superalgebra (6) allows for a complete
group-theoretical solution of the Pauli equation, along the lines
indicated by D'Hoker and Vinet [2]. Details will be given elsewhere.

Notice that Jackiw's two-dimensional Dirac operator $\pib\times\sigma$
in Ref. [8]
 --- essentially our $Q^\star$ --- is
associated with the unusual choice of the two-dimensional `Dirac'
(i.e. Pauli)
matrices
$\gamma_1^\star\=-\sigma_2$,
$\gamma_2^\star\=\sigma_1$. 
Our helicity operator, $Q$,
is again a \lq Dirac operator' --- but one associated with the standard choice
$\gamma_1\=\sigma_1$,
$\gamma_2\=\sigma_2$.

\kikezd{Note added}.
After this paper was completed and even submitted, we became aware of some papers [11] which expressed similar ideas. Our paper has 
consequently remained unpublished; parts of it entered [12].

\vskip2mm

\centerline{\bf References}

\reference
R.~Jackiw,
Ann.~Phys. (N.Y.) {\bf 129}, 183 (1980).

\reference E.~D'Hoker and L.~Vinet,
Phys.~Lett. {\bf 137B}, 72 (1984);
Comm.~Math.~Phys. {\bf 97}, 391 (1985).

\reference
R.~Jackiw,
Ann.~Phys. (N.Y.) {\bf 201}, 83 (1990).

\reference
Y.~Aharonov and D.~Bohm,
Phys.~Rev. {\bf 115}, 485 (1959).

\reference
E.~Witten,
Nucl.~Phys. {\bf B185}, 513 (1981);
P.~Salomonson and J.W.~Van Holten,
Nucl.~Phys. {\bf B169}, 509 (1982);
M.~De Crombrugghe and V.~Rittenberg, Ann.~Phys. (N.Y.)
{\bf 151}, 99 (1983).

\reference
M.~Leblanc, G.~Lozano and H.~Min,
Ann. Phys. (N.Y.) {\bf 219}, 328 (1992);
C.~Duval and P.A.~Horv\'athy,
submitted to Comm.~Math.~Phys.
The bosonic galilean symmetry was noticed by
R. Jackiw and S.-Y. Pi, Phys. Rev. {\bf D 42}, 3500 (1990).

\reference
R. J. Hughes, V. A. Kosteleck\'y and M. M. Nieto,
Phys. Rev. {\bf D34}, 1100 (1986);
E. D'Hoker, V. A. Kosteleck\'y and L. Vinet, in
{\it Dynamical groups and spectrum generating algebras},
A. Bohm, Y. Ne'eman and A. O. Barut (eds), Vol. 1, p. 339;
Singapore: World Scientific (1988).

\reference
R.~Jackiw,
Phys.~Rev. {\bf D29}, 2375 (1984).

\reference
C. R. Hagen, Phys. Rev. Lett. {\bf 64}, 503 (1990);
R.~Musto, L.~O'Raifeartaigh and A.~Wipf,
Phys. Lett. {\bf B 175}, 433 (1986);
P. Forg\'acs, L.~O'Raifeartaigh and A.~Wipf,
Nucl. Phys. {\bf B 293}, 559 (1987).

\reference
P.A.~Horv\'athy, Phys.~Rev. {\bf A31}, 1151 (1985);
F.~Wilczek, Phys.~Rev.~Lett. {\bf 48}, 1144 (1982);
R.~Jackiw and A.N.~Redlich, Phys.~Rev.~Lett. {\bf 50}, 555 (1983)
W.C.~Henneberger, Phys. Rev. Lett. {\bf 52}, 573 (1984).

\reference
C. J. Park,
Nucl. Phys. {\bf 376}, 99 (1992);
J.-G. Demers,
Mod. Phys. Lett. {\bf 8}, 827 (1993).

\reference
C. Duval and P. A. Horv\'athy, J. Math. Phys. {\bf 35}, 2516 (1994)
[hep-th/0508079]
\bye